\documentclass[12pt,preprint]{aastex}

\slugcomment{Draft \today; for submission to ApJL}

\newcommand{\skipthis}[1]{}

\newcommand{\kms}{\hbox{km\,s$^{-1}$}}

\shortauthors{Gutermuth et al.}
\shorttitle{Embedded Cluster Discovered in Serpens-Aquila Rift}

\begin{document}

\title{The {\it Spitzer} Gould Belt Survey of Large Nearby Interstellar Clouds:
Discovery of a Dense Embedded Cluster in the Serpens-Aquila Rift}

\author{R. A. Gutermuth\altaffilmark{1},
T. L. Bourke\altaffilmark{1}, 
L. E. Allen\altaffilmark{1}, 
P. C. Myers\altaffilmark{1},
S. T. Megeath\altaffilmark{2},
B. C. Matthews\altaffilmark{3},
J. K. J{\o}rgensen\altaffilmark{4},
J. Di Francesco\altaffilmark{3},
D. Ward-Thompson\altaffilmark{5},
T. L. Huard\altaffilmark{1},
T. Y. Brooke\altaffilmark{6},
M. M. Dunham\altaffilmark{7},
L. A. Cieza\altaffilmark{7},
P. M. Harvey\altaffilmark{7},
N. L. Chapman\altaffilmark{8}
}

\altaffiltext{1}{Harvard-Smithsonian Center for Astrophysics, 60 Garden
Street, Cambridge, MA 02138; email rgutermuth@cfa.harvard.edu}
\altaffiltext{2}{Ritter Observatory, Department of Physics and Astronomy, University of Toledo, Toledo, OH 43606}
\altaffiltext{3}{Herzberg Institute of Astrophysics, National Research Council of Canada, Victoria, BC V9E 2E7, Canada}
\altaffiltext{4}{Argelander-Institut f\"{u}r Astronomie, University of Bonn, Auf dem H\"{u}gel 71, 53121 Bonn, Germany}
\altaffiltext{5}{Department of Physics and Astronomy, University of Cardiff, P.O. Box 913, Cardiff, CF2 3YB, Wales, UK}
\altaffiltext{6}{Spitzer Science Center, MC 220-6, California Institute of Technology, Pasadena, CA 91125}
\altaffiltext{7}{Department of Astronomy, University of Texas, Austin, TX 78712}
\altaffiltext{8}{Department of Astronomy, University of Maryland, College Park, MD 20742}

\begin{abstract}

We report the discovery of a nearby, embedded cluster of young stellar objects,
associated filamentary infrared dark cloud, and 4.5~$\mu$m shock emission knots
from outflows detected in {\it Spitzer}/IRAC mid-infrared imaging of the
Serpens-Aquila Rift obtained as part of the {\it Spitzer} Gould Belt Legacy
Survey.  We also present radial velocity measurements of the region from
molecular line observations obtained with the Submillimeter Array (SMA) that
suggest the cluster is co-moving with the Serpens Main embedded cluster
$3^\circ$ to the north.  We therefore assign it the same distance, 260~pc.  The
core of the new cluster, which we call Serpens South, is composed of an
unusually large fraction of protostars (77\%) at high mean surface density
($>$430~pc$^{-2}$) and short median nearest neighbor spacing (3700~AU).  We
perform basic cluster structure characterization using nearest neighbor surface
density mapping of the YSOs and compare our findings to other known clusters
with equivalent analyses available in the literature.  

\end{abstract}

\keywords{ISM: individual (IRAS 18275-0203, IRAS 18274-0205) -- stars:
formation -- stars: low-mass} 

\section{Introduction}

The aim of the {\it Spitzer} Gould Belt (GB; Allen et al. in prep.) and c2d
\citep{evan03} Legacy Surveys is to build a complete mid-infrared record of
star formation in all of the large molecular clouds within 500~pc, largely
dominated by the Gould Belt itself.  This is one of several surveys of the
Gould Belt being performed at many wavelengths from the near-IR
\citep[Two-Micron All Sky Survey; 2MASS][]{skru06} through the far-infrared
\citep[{\it Herschel};][]{as05} to the submillimeter \citep[Submillimeter
Common User Bolometer Array 2; SCUBA-2;][]{ward07}.  In combination, the
resulting database offers a nearly {\it complete and unbiased} view of all
nearby star formation in large molecular clouds.  This penetrating examination
of regions that have often received little attention in the literature is
likely to reveal new regions of star formation.

The {\it Spitzer} Infrared Array Camera (IRAC) imaging presented here reveals a
dense cluster of sources with excess infrared emission and associated
4.5~$\mu$m-bright structured nebulosity \citep[typically shocked H$_2$ emission
from outflows, cf.][]{smit06} within a filamentary infrared dark cloud seen in
absorption against bright diffuse 5.8 and 8.0~$\mu$m emission.  This emission
is most likely PAH-feature emission \citep{morr04} excited by the young, high
mass stars in W40 \citep{smit85}, 20\arcmin\ to the east.  We found 
structured $K_S$ nebulosity at the site in 2MASS Atlas images, but few sources.
The region is flanked by two {\it Infrared Astronomy Satellite} (IRAS) sources
(IRAS~18275-0203 and IRAS~18274-0205), and {\it Midcourse Space Experiment}
(MSX) coverage of this region reveals only a few additional point sources.  No
deep X-ray images of this region are available.

\section{Observations}


We obtained 3.6, 4.5, 5.8, and 8.0~$\mu$m imaging of 5.1 square degrees of the
Serpens-Aquila Rift with IRAC \citep{fazi04} onboard {\it Spitzer}
\citep{wern04} as part of the GB Legacy Survey (PID: 30574) on 27 October 2006.
In this letter, we focus on the 14\arcmin~$\times$~10\arcmin\ field of view
centered on Serpens South (R.A.,~Dec.~(2000) = 18:30:03, $-$02:01:58.2).  High
Dynamic Range mode was used, resulting in 0.4 and 10.4 second integrations for
each of four dithered images at each mosaic position.  Mosaics were constructed
at 1\farcs2 per pixel with Basic Calibrated Data (BCD) products from the {\it
Spitzer} Science Center's data pipeline v14.4.
All artifact treatment, mosaic construction, point source detection and
aperture photometry, and photometric classification of sources presented here
has been performed using custom Interactive Data Language (IDL) routines
\citep{gute07}. 
We adopted aperture radii and inner and outer background annulus radii of
2\farcs4, 2\farcs4, and 7\farcs2, respectively.  Point source flux calibration
is discussed in detail in \citet{reac05}.  Field averaged 90\% differential
completeness magnitude limits are 14.5, 14.2, 13.2, and 12.0 for the
14\arcmin~$\times$~10\arcmin\ submosaics at 3.6, 4.5, 5.8, and 8.0~$\mu$m,
respectively \citep{gute05}.  The sensitivity of these data are equivalent to
the IRAC data presented by \citet{wins07} for the Serpens Main cluster, which
will be discussed in Section~\ref{disc}.  
IRAC photometry catalogs are bandmerged together with a radial matching
tolerance of 1\arcsec, and then merged with the 2MASS Point Source Catalog
(PSC) at a tolerance of 1\arcsec.

\subsection{Submillimeter Array}

Observations with the SMA\footnote{The Submillimeter Array is a joint project
between the Smithsonian Astrophysical Observatory and the Academia Sinica
Institute of Astronomy and Astrophysics and is funded by the Smithsonian
Institution and the Academia Sinica.} \citep{ho04} were performed on 30 January
2007.  The array was tuned to cover the frequency range 219.4-221.4 GHz (LSB)
and 229.4-231.4 GHz (USB), which includes the lines of $^{12}$CO, $^{13}$CO,
and C$^{18}$O 2-1, with a resolution of $\sim$0.5 \kms/channel.  Three
positions were observed for 30 minutes each; IRAS 18275-0203 (R.A.,Dec.~(2000)
= 18:30:05.8, $-$02:01:45) and a nearby embedded source (R.A.,Dec.~(2000) =
18:30:01.3, $-$02:01:48, labelled ``Embedded'' in Figure~\ref{pretty1}), both
in Serpens South, and a position near to the $^{12}$CO 1-0 peak of W40, as
determined from the maps of \citet{grab87}, at R.A.,Dec.~(2000) = 18:31:20.7,
$-$02:01:07.  Ganymede was observed for passband calibration.  No flux or
complex gain calibration was performed.  More detailed SMA results will be
discussed in a future paper.

\section{Results}

\subsection{Distance\label{dist}}

Serpens South appears to be associated with a filamentary cloud seen in
absorption against PAH emission from W40 (Fig.~\ref{pretty1}).  Thus the
distance to W40 \citep[550~$\pm$~150~pc][]{radh72,smit85} is an upper limit to
the distance to Serpens South.  The Serpens Main cluster has local standard of
rest (LSR) velocities in the range 6-11 \kms, when observed in the
isotopologues of CO \citep{whit95}, with Gaussian FWHM widths of $\sim2$ \kms.
Thus, its molecular line emission can be separated from that associated with
W40, which occurs in the LSR range of 2-6 \kms\ \citep{zl78,vall92,zhu06}, and
is spatially compact.  The SMA data clearly show that the velocities of IRAS
18275-0203 and the nearby embedded source are the same as that of the Serpens
Main cluster, and not of the W40 cluster.  We therefore assign the distance of
260~$\pm$~37~pc \citep{stra96} to Serpens South.  This is consistent with more
recent results for the entire Serpens-Aquila Rift \citep{stra03}.

\subsection{Identification, Membership, and Spatial Distribution of YSOs\label{id}}

\begin{figure*}
\epsscale{1}
\plotone{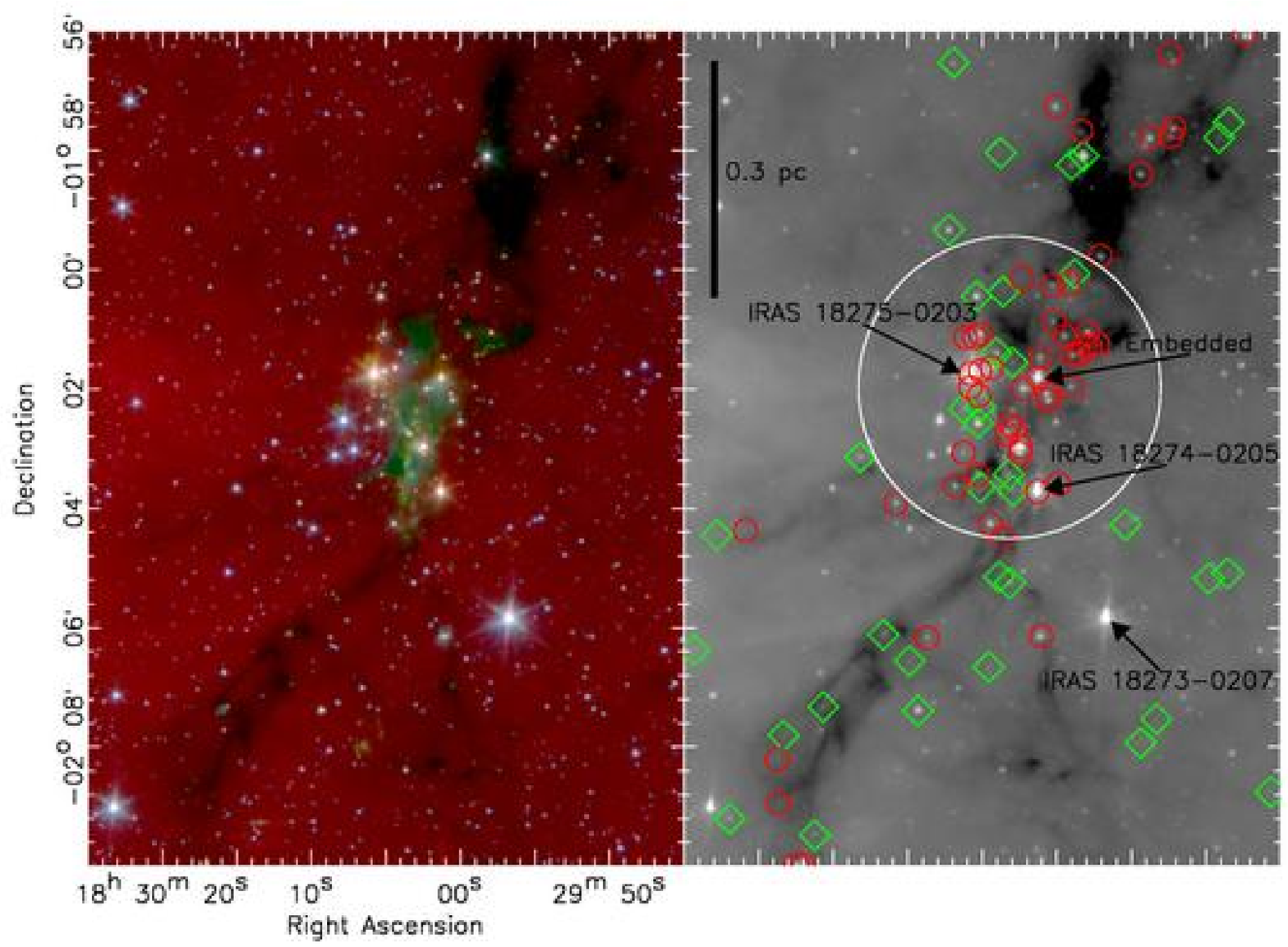}
\caption{At left, a color-composite image of the IRAC mosaics of Serpens South; 3.6, 4.5, and 8.0~$\mu$m images are mapped to blue, green, and red, respectively. At right, the spatial distribution of YSOs overlaid on the grayscale 8.0~$\mu$m image.  Red circles are Class~I protostars, and green diamonds are Class~II stars with disks.  The white circle marks the dense core of the cluster.  
\label{pretty1}}
\end{figure*}

We identify likely YSOs from their excess infrared emission over that expected
from a typical photosphere and high infrared flux compared to extragalactic
contaminants \citep{gute07}.  These candidates are classified as Class~I
(protostar) or Class~II (pre-main sequence star with disk) YSOs\footnote{Class
III (diskless) YSOs cannot be distinguished from field stars in this way
\citep{mege04}, but can be effectively identified with X-ray observations
\citep[cf.][]{feig07}.} by the their value of $\alpha_{IRAC}$, the linear least
squares fit to their spectral energy distribution (log~$\lambda
F_\lambda$~vs.~log~$\lambda$) through the four IRAC bandpasses \citep{lada06}.
In this way, we have identified 54~Class~I and flat spectrum ``protostars''
\citep[$\alpha_{IRAC}>-0.3$][]{gree94} and 37~Class~II ($-1.6 \le \alpha_{IRAC}
\le -0.3$) YSOs, a total of 91 (59\% protostars), within the
14\arcmin~$\times$~10\arcmin\ (1.1~pc~$\times$~0.8~pc) field of view presented
in Fig.~\ref{pretty1}.  
Half of the sources in this field of view are located in a region that stands
out as both relatively high in surface density and dominated by sources that
are protostellar.  Of the YSOs identified, 37 Class~I and 11 Class~II sources
are located within a 2\farcm5 (0.2~pc) radius circle centered on
R.A.,~Dec.~(2000) = 18:30:03, $-$02:01:58.2, making up the core of the Serpens
South Cluster (Fig.~\ref{pretty1}).  In this core, 77\% of the YSOs are
protostellar, with a mean density of 430~pc$^{-2}$.  This dense grouping
appears elongated in an alignment similar to the dust filament seen in
absorption (150$^\circ$ east of north).  The asymmetric distribution suggests
that the mean surface density of 430~pc$^{-2}$ in the core is likely a
significant underestimate.  To demonstrate this, we note that the median
projected distance between nearest neighbor YSOs here is 13\farcs2, or 3700~AU.
Sources uniformly spaced at a density of 430~pc$^{-2}$ are 21\farcs4 apart, or
5600~AU.

\subsection{YSO Surface Density Mapping\label{nnmap}}

\begin{figure}
\epsscale{.7}
\plotone{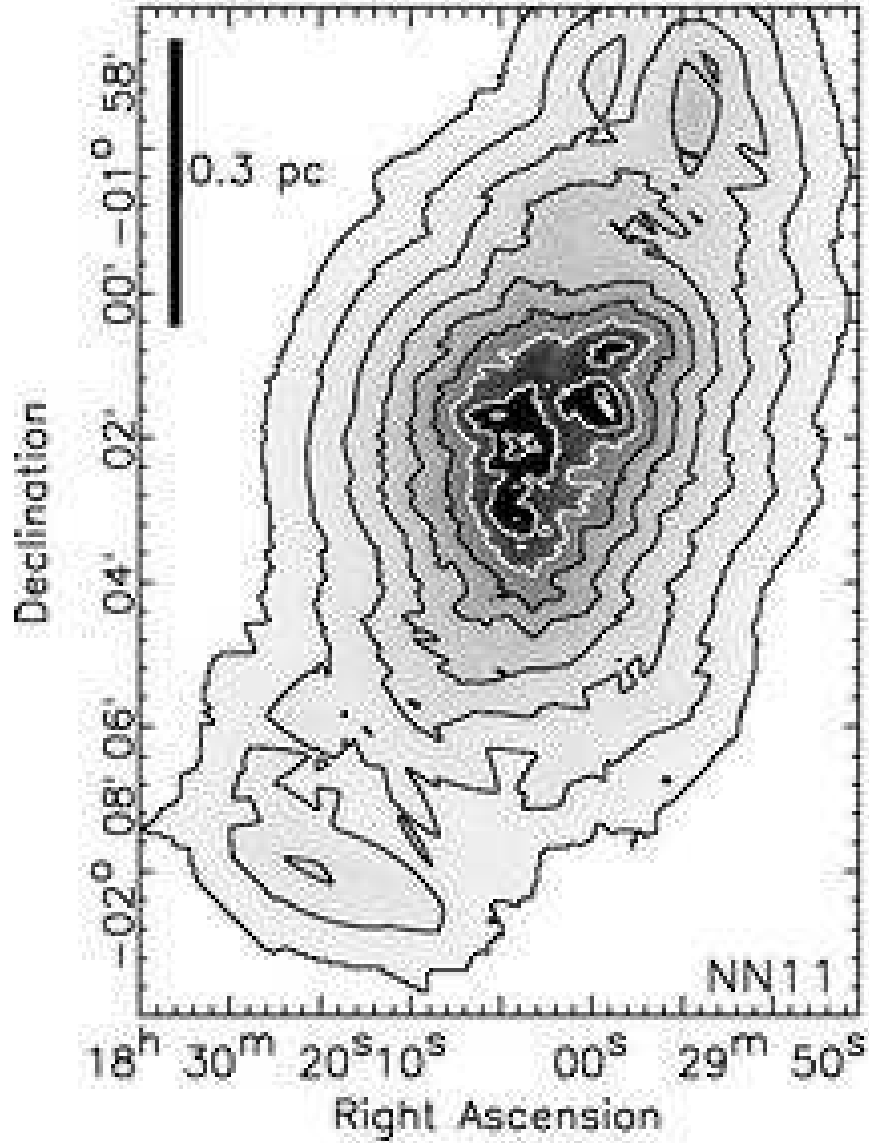}
\caption{A $n=11$ nearest neighbor surface density map for the Serpens South YSOs.  Contour levels mark stellar surface densities of 1~$\sigma$ (33\%) below successive contours (1300, 890, \& 590 pc$^{-2}$ in white and 400, 260, 180, 78, \& 52 pc$^{-2}$ in black), and the linear inverse grayscale sets 1000~pc$^{-2}$ as black and 0~pc$^{-2}$ as white.  The field of view is the same as for Fig.~\ref{pretty1}. 
\label{surf1}}
\end{figure}

To characterize the two dimensional structure of the Serpens South cluster, we
have constructed a nearest neighbor surface density map of all the {\it
Spitzer}-identified YSOs (Fig.~\ref{surf1}).  The method used to generate these
maps is documented in \citet{gute05}\footnote{In summary, from each position in
a uniform grid, we measure the distance $r_n(i,j)$ that defines a circle that
contains the nearest $n$ sources.  From this radial distance, a surface density
is directly computed as $\sigma_n(i,j) = \frac{n-1}{\pi r_n^2(i,j)}$
\citep{ch85}.  
}, though we have used $n=11$ nearest neighbors here to sample the surface
densities at higher fidelity \citep[33\% uncertainty;][]{ch85}.  The elevated
density is evident in the center of the map, with a peak density of
1600~pc$^{-2}$.  The apparent boundary contour for the dense cluster core is
590~pc$^{-2}$, a more reasonable mean surface density than was measured over
the poorly matched circular area used in Section~\ref{id} above.  Lower density
star formation along the dust filaments is also apparent at densities from 50
to 120~pc$^{-2}$.  The overall structure is elongated, reflecting the same
orientation and extent as the filamentary cloud structures seen in absorption
(Fig.~\ref{pretty1}).

\section{Putting Serpens South in Context\label{disc}}

\begin{figure}
\epsscale{.5}
\plotone{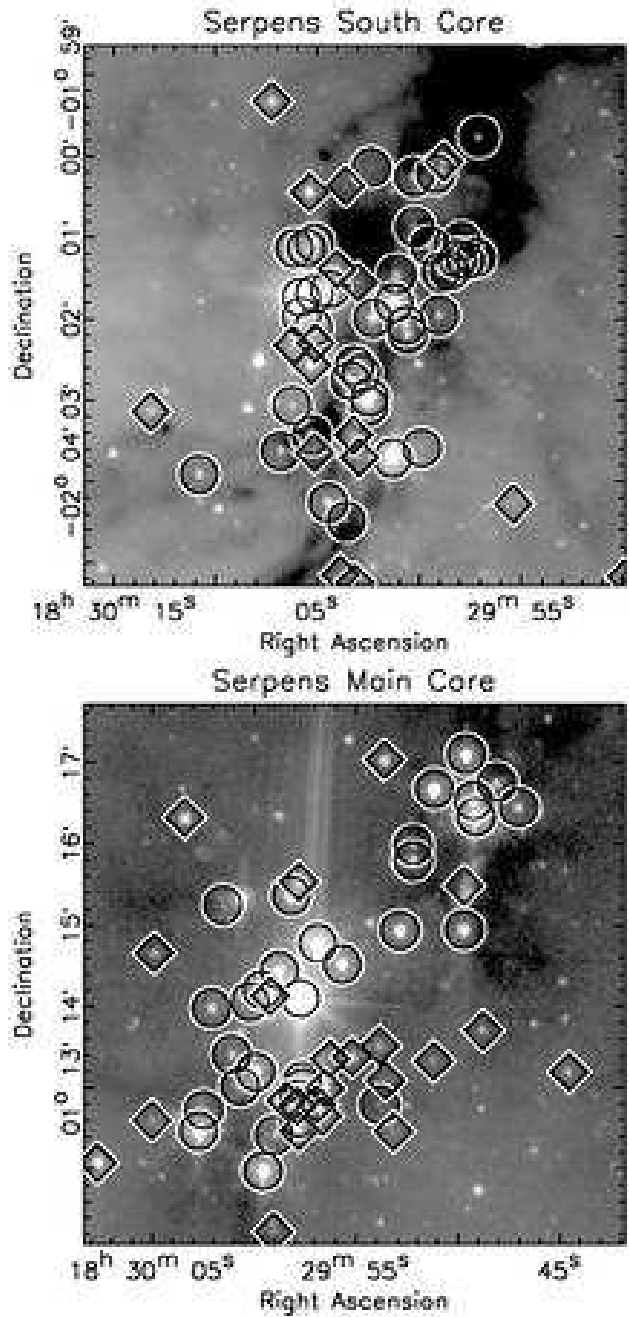}
\caption{{\it Spitzer}/IRAC 8~$\mu$m images of the 0.5~pc~$\times$~0.5~pc core regions of the Serpens South (this work) and Serpens Main \citep{wins07} clusters.  Overlaid are the positions of the Class~I sources (circles) and the Class~II sources (diamonds) for each region.  We have included the flat spectrum sources from \citet{wins07} as part of the Class~I population for the purposes of this plot. 
\label{sbs1}}
\end{figure}

The Serpens South 
core's high mean surface density ($>$430~pc$^{-2}$), exceptionally large
fraction of Class~I sources (77\%), and relatively large membership (48, with
43 more along the 0.5~pc cloud filaments) suggest both a very recent onset of
star formation \citep[within the typical lifetime of the protostellar phase, $2
\pm 1 \times 10^5$~yr, cf.][]{kh95} and high star formation rate
($\sim$90~$M_\odot$~Myr$^{-1}$ in the core, assuming the above protostellar
phase lifetime and 0.5~$M_\odot$ per source).  Visually, the region bears some
resemblance to the Serpens Main cluster \citep{harv06,wins07} that we have
shown lie at similar distances.  In Fig.~\ref{sbs1}, we present
6.5\arcmin~$\times$~6.5\arcmin\ (0.5~pc~$\times$~0.5~pc) 8~$\mu$m images of the
dense cores of both the Serpens South and Serpens Main clusters, overplotted
with the positions of {\it Spitzer} identified Class~I and Class~II sources
\citep{wins07}.  Over these fields of view, the two regions have nearly
identical numbers of YSOs with excess infrared emission and similar protostar
fractions (55 at 71\% and 54 at 57\% in Serpens South and Serpens Main,
respectively).  However, Serpens South is more concentrated and confined to its
filament axis than Serpens Main, as evidenced by median nearest neighbor
distances between YSOs of 3700~AU and 4800~AU, respectively.
These results suggest that Serpens South is at least as young and as prolific a
star forming region as Serpens Main, though sampling statistics prevent us from
making any claims beyond similarity.  At larger scales, both Serpens Main and
Serpens South are active star forming sites embedded in larger, less dense,
less extinguished distributions of YSOs.  Given the high likelihood that both
clusters are associated with the Serpens-Aquila Rift and their structural and
evolutionary similarity over such a large projected distance from each other
(197\arcmin\ or 15~pc), we speculate that the initial conditions of this region
play a strong role in defining these qualities.

Serpens South joins a growing number of young clusters that are protostar rich
and asymmetrically structured.
The more typical asymmetric clusters (e.g. NGC~1333, IRAS~20050+2720,
GGD~12-15) have evolved to the point that star formation is active throughout
most sites of dense gas in their natal clouds.  This is supported by their
large numbers of protostars (active current star formation), their larger
numbers of stars with disks (star formation in the recent past), and the
distributions of both YSO types that appear to reflect the typically
filamentary dense gas distributions \citep{gt05,ppv07}.  Serpens South has a
particularly high protostar fraction though \citep[59\%; NGC 1333 is 29\%
protostars, excluding transition disks;][]{gute07}, and it is part of a dense,
dusty cloud filament that is currently forming stars at a rather low surface
density 
compared to the core.  If the rest of the filament continues forming stars
within the next few $10^5$~yr, we speculate that this region could evolve into
an asymmetric cluster of similar number of sources and protostar fraction to
the aforementioned asymmetric examples.  Therefore, the Serpens South cluster
could be described as a {\it protocluster}, a region with a sufficiently high
star formation rate to produce a large number of stars before gas dispersal
processes (outflows, radiation, etc.) from the growing
cluster membership terminate the process.  Given its close proximity, follow-up
observations (e.g. to characterize the gas distribution and kinematics) and
detailed analysis should offer a new and unique view of the earliest stages of
clustered star formation. 

\acknowledgements

We thank T. Dame and E. Winston for providing data from \citet{grab87} and \citet{wins07} in electronic form.
This publication makes use of data products from the Two Micron All Sky Survey,
which is a joint project of the University of Massachusetts and the Infrared Processing and Analysis Center/California Institute of Technology, funded by the National Aeronautics and Space Administration and the National Science Foundation.This research has made use of the SIMBAD database, operated at CDS, Strasbourg,
France.
This work is based in part on observations made with the Spitzer Space Telescope, which is operated by the Jet Propulsion Laboratory, California Institute of Technology under a contract with NASA. 


\begin{thebibliography}{}

\bibitem[Allen et al.(2007)]{ppv07} Allen, L., et al.\ 2007, Protostars and Planets V, 361 

\bibitem[Andr{\'e} \& Saraceno(2005)]{as05} Andr{\'e}, P., \& Saraceno, P.\ 2005, ESA Special Publication, 577, 179

\bibitem[Casertano \& Hut(1985)]{ch85} Casertano, S., \& Hut, P.\ 1985, \apj, 298, 80

\bibitem[Dame et al.(2001)]{dame01} Dame, T.~M., Hartmann, D., \& Thaddeus, P.\ 2001, \apj, 547, 792 

\bibitem[Evans et al.(2003)]{evan03} Evans, N.~J., II, et al.\ 
2003, \pasp, 115, 965 

\bibitem[Fazio et al.(2004)]{fazi04} Fazio, G. G. et al. 2004, \apjs, 154, 10

\bibitem[Feigelson et al.(2007)]{feig07} Feigelson, E., Townsley, L., G{\"u}del, M., \& Stassun, K.\ 2007, Protostars and Planets V, 313

\bibitem[Flaherty et al.(2007)]{flah07} Flaherty, K.~M., Pipher, J.~L., Megeath, S.~T., Winston, E.~M., Gutermuth, R.~A., Muzerolle, J., Allen, L.~E., \& Fazio, G.~G.\ 2007, \apj, 663, 1069 

\bibitem[Grabelsky et al.(1987)]{grab87} Grabelsky, D.~A., Cohen, R.~S., Bronfman, L., Thaddeus, P., \& May, J.\ 1987, \apj, 315, 122 

\bibitem[Greene et al.(1994)]{gree94} Greene, T.~P., Wilking, B.~A., Andr{\'e}, P., Young, E.~T., \& Lada, C.~J.\ 1994, \apj, 434, 614 

\bibitem[Gutermuth et al.(2005)]{gute05} Gutermuth, R. A., Megeath, S.T., Pipher, J. L., Williams, J. P., Allen, L. E., Myers, P. C., \& Raines, S. N.\ 2005, \apj, 632, 397

\bibitem[Gutermuth (2005)]{gt05} Gutermuth, R. A.\ 2005, PhD thesis, University of Rochester

\bibitem[Gutermuth et al.(2007)]{gute07} Gutermuth, R.~A., et al.\ 2007, ArXiv e-prints, 710, arXiv:0710.1860 

\bibitem[Harvey et al.(2006)]{harv06} Harvey, P.~M., et al.\ 2006, \apj, 644, 307

\bibitem[Ho et al.(2004)]{ho04} Ho, P.~T.~P., Moran, J.~M., \& Lo, K.~Y.\ 2004, \apjl, 616, L1 

\bibitem[Kenyon \& Hartmann(1995)]{kh95} Kenyon, S.~J., \& Hartmann, L.\ 1995, \apjs, 101, 117 

\bibitem[Lada et al.(2006)]{lada06} Lada, C.~J., et al.\ 2006, \aj, 131, 1574

\bibitem[Megeath et al.(2004)]{mege04} Megeath, S.~T., et al.\ 2004, \apjs, 154, 367

\bibitem[Morris et al.(2004)]{morr04} Morris, P.~W., Crowther, P.~A., \& Houck, J.~R.\ 2004, \apjs, 154, 413 

\bibitem[Radhakrishnan et al.(1972)]{radh72} Radhakrishnan, V., Goss, W.M., Murray, J.D., \& Brooks, J.W.\ 1972, \apjs, 24, 49

\bibitem[Reach et al.(2005)]{reac05} Reach, W.~T., et al.\ 2005, \pasp, 117, 978

\bibitem[Ridge et al.(2003)]{ridg03} Ridge, N. A.. Wilson, T. L., Megeath, S. T., Allen, L. E., Myers, P. C., 2003, \aj, 126, 286

\bibitem[Skrutskie et al.(2006)]{skru06} Skrutskie, M.~F., et al.\ 2006, \aj, 131, 1163 

\bibitem[Smith et al.(1985)]{smit85} Smith, J., Bentley, A., Castelaz, M., Gehrz, R. D., Grasdalen, G. L., \& Hackwell, J. A.\ 1985, \apj, 291, 571

\bibitem[Smith et al.(2006)]{smit06} Smith, H.~A., Hora, J.~L., Marengo, M., \& Pipher, J.~L.\ 2006, \apj, 645, 1264 

\bibitem[Strai{\v z}ys et al.(1996)]{stra96} Strai{\v z}ys, V., {\v C}ernis, K., \& Barta{\v s}i{\= u}t{\.e}, S.\ 1996, Baltic Astronomy, 5, 125 

\bibitem[Strai{\v z}ys et al.(2003)]{stra03} Strai{\v z}ys, V., {\v C}ernis, K., \& Barta{\v s}i{\= u}t{\.e}, S.\ 2003, \aap, 405, 585 

\bibitem[Vallee et al.(1992)]{vall92} Vallee, J.~P., Guilloteau, S., \& MacLeod, J.~M.\ 1992, \aap, 266, 520 

\bibitem[Ward-Thompson et al.(2007)]{ward07} Ward-Thompson, D., et al.\ 2007, \pasp, 119, 855

\bibitem[Werner et al.(2004)]{wern04} Werner, M.~W., et al.\ 2004, \apjs, 154, 1

\bibitem[White et al.(1995)]{whit95} White, G.~J., Casali, M.~M., \& Eiroa, C.\ 1995, \aap, 298, 594 

\bibitem[Winston et al.(2007)]{wins07} Winston, E., et al.\ 2007, \apj, 669, 493

\bibitem[Zeilik \& Lada(1978)]{zl78} Zeilik, M., II, \& Lada, C.~J.\ 1978, \apj, 222, 896 

\bibitem[Zhu et al.(2006)]{zhu06} Zhu, L., Wu, Y.-F., \& Wei, Y.\ 2006, Chinese Journal of Astronomy and Astrophysics, 6, 61 

\end{thebibliography}
\end{document}